\documentclass[a4paper,10pt]{article}

\usepackage[utf8x]{inputenc}
\usepackage{fullpage}
\usepackage{graphicx}
\usepackage{url}
\usepackage{algorithm}
\usepackage{algorithmic}
\usepackage{color}
\usepackage{amssymb}
\usepackage[table]{xcolor}
\usepackage{amsmath}
\usepackage{amsthm}
\usepackage{xyling}
\usepackage{subfigure}
\usepackage{authblk}

\newcommand\cardl{\left|}
\newcommand\cardr{\right|}

\newtheorem{theorem}{Theorem}[section]
\newtheorem{lemma}[theorem]{Lemma}

\begin{document}
\title{Secure Data Processing in a Hybrid Cloud}

\author{Vaibhav Khadilkar, Murat Kantarcioglu, Bhavani Thuraisingham \\ The University of Texas at Dallas \\ \texttt{\{vvk072000, muratk, bxt043000\}@utdallas.edu} \and Sharad Mehrotra \\ University of California, Irvine \\ \texttt{sharad@ics.uci.edu}}
\date{}
\maketitle

\begin{abstract}
Cloud computing has made it possible for a user to be able to select a computing service precisely when needed. However, certain factors such as security of data and regulatory issues will impact a user's choice of using such a service. A solution to these problems is the use of a hybrid cloud that combines a user's local computing capabilities (for mission- or organization-critical tasks) with a public cloud (for less influential tasks). We foresee three challenges that must be overcome before the adoption of a hybrid cloud approach: 1) \textit{data design}: How to partition relations in a hybrid cloud? The solution to this problem must account for the sensitivity of attributes in a relation as well as the workload of a user; 2) \textit{data security}: How to protect a user's data in a public cloud with encryption while enabling query processing over this encrypted data? and 3) \textit{query processing}: How to execute queries efficiently over both, encrypted and unencrypted data? This paper addresses these challenges and incorporates their solutions into an add-on tool for a Hadoop and Hive based cloud computing infrastructure.
\end{abstract}
\section{Introduction}
\label{sec:intro}
The emergence of cloud computing has created a paradigm shift by allowing parallel processing of massive amounts of data. Cloud computing has further segmented traditionally provided software services into SaaS, PaaS and IaaS. This segmentation allows users to choose the appropriate kind of computing service precisely when needed. Further, using cloud computing services can significantly lower a user's capital expenditure since they only pay for services that they use. However, a user needs to make an informed decision as to whether or not to use cloud services based on other factors such as the level of information privacy desired, regulatory issues and local computing capacity. A user may be tempted to use other secure data processing alternatives such as full homomorphic encryption \cite{homenc}. Unfortunately such methods are very expensive as the the data size increases. Given these issues, for certain users it may be a better choice to adopt a hybrid cloud (public and private) approach rather than relying solely on a cloud service provider. Further, this hybrid solution enables certain mission- or organization-critical tasks to be executed locally at a user's site while allowing less important tasks to be outsourced to the public cloud. Moreover, this increases throughput while reducing operational costs with a high-level of data security. 

There are a number of technological issues that need to be addressed before the adoption of a hybrid cloud methodology. The first issue is, how to distribute data in a hybrid cloud? This is the \textit{data design} problem which focuses on how data should be partitioned and where these partitions should be placed. The main reasons for data distribution are scalability, higher concurrency and greater throughput. There are a number of related design issues such as granularity of partitions and application requirements. Data design, especially in the cloud computing paradigm is a challenging task. This is because certain attributes of a user's data may be sensitive, in which case the user cannot release this information to a cloud service provider unless it is encrypted. Our data design module takes into account this factor during the process of data fragmentation and fragment allocation. 

The next issue is, how do users protect themselves from cloud service providers who may be able to access their data? This issue is related to \textit{data security} and is relevant for users since their data is placed at the provider's site. The goal of data security is to prevent the service provider from learning any meaningful information from the data. Data security can be achieved by encrypting the stored data. However, encryption presents a new set of challenges such as granularity of encryption and query processing over encrypted data. Our security module addresses these challenges.
%
\begin{figure*}[htp]
\centering
\includegraphics[width=155mm,height=95mm]{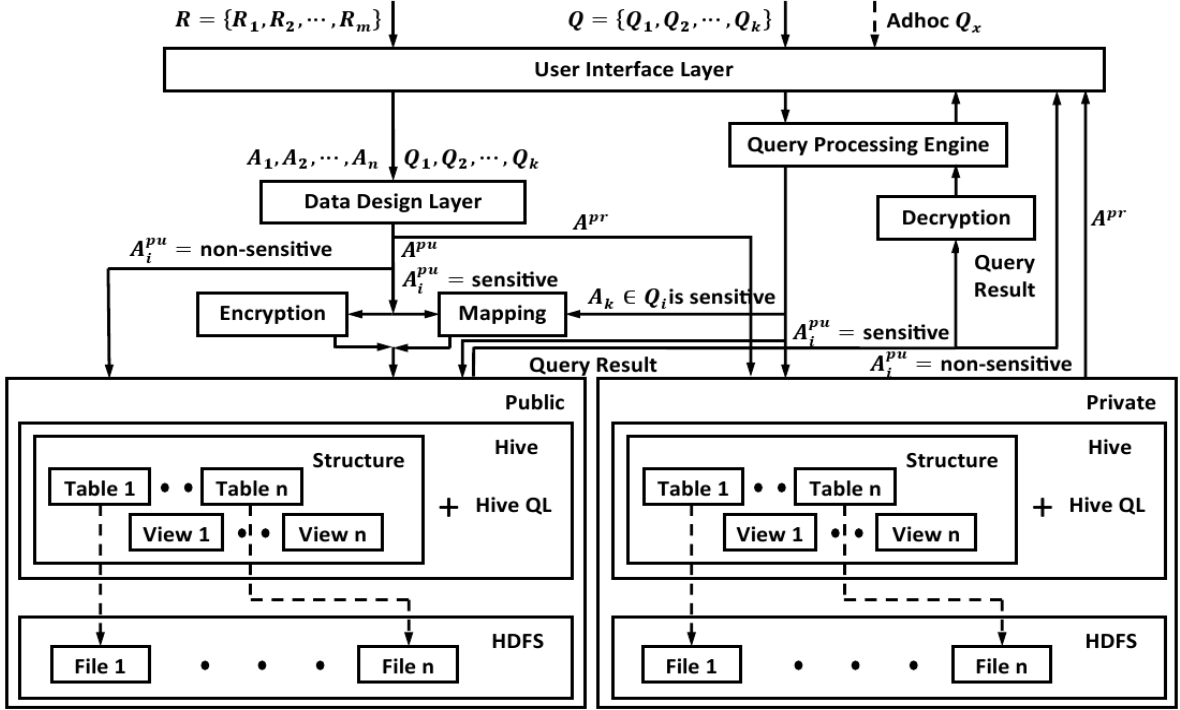}
\hfill
\caption{The hybrid cloud architecture}
\label{fig:architecture}
\end{figure*} 

The final issue is, how can users query the cloud infrastructure without being aware of the separation of data in a hybrid cloud? This is the problem of \textit{query processing} in a distributed environment. The goal of a cloud query processor is to transform a high-level query into a low-level query that can be executed over a hybrid cloud. This query processor must be able to take into account the cost of executing queries over a hybrid cloud containing both, unencrypted and encrypted data. Further, this query processor must be able to optimize query execution given the fragmentation of relations as well as the sensitivity of attributes. Our query processing module solves these problems effectively. 

Figure~\ref{fig:architecture} presents the architecture of our proposed system. A user submits the original set of relations, $R = \{R_{1}, R_{2}, \cdots, R_{m}\}$ and input queries, $Q = \{Q_{1}, Q_{2}, \ldots, Q_{k}\}$. The \textit{data design} layer vertically partitions the set of attributes $A = \{A_{1}, A_{2}, \ldots, A_{n}\}$ over all relations $R$ into $A^{pu}$ and $A^{pr}$ which are the sets of attributes stored on the public and private clouds respectively. These sets are determined by solving an optimization problem that minimizes the cost of executing $Q$ over $A^{pu}$ and $A^{pr}$. Then, the sensitive attributes in $A^{pu}$ are encrypted and mapped before being stored while the non-sensitive attributes are stored unencrypted in $A^{pu}$ and $A^{pr}$. The \textit{query processing engine} takes a query $q_{i} \in Q$ or an ad-hoc query, $q_{x}$, and transforms it into a query over a hybrid cloud. The results obtained by executing queries on both clouds are then combined into the final result that is passed on to a user. We use a Hadoop HDFS and Hive based cloud storage infrastructure for our implementation. Hadoop HDFS is a distributed file system that is designed to run on commodity hardware. In our architecture, the HDFS layer stores files that contain the vertical partitions created in the data design phase. A file may contain unencrypted or encrypted data depending on the sensitivity of attributes stored in that file. Hive is a data warehouse that is built on Hadoop. Hive allows a user to define structure for files that are stored in the underlying HDFS. Furthermore, Hive provides a user the ability to query this structured data using a SQL-like query language called Hive QL. In our architecture, Hive is used to create the public and private components of a relation using the vertical partitions stored in HDFS. Then, a query $q$ is split into two sets of sub-queries using Hive QL: $q^{pu}$ is executed on the public cloud, while $q^{pr}$ is executed on the private cloud. The results of $q^{pu}$ and $q^{pr}$ are combined on the private cloud and then returned to a user.

\textbf{Our contributions:} To address the challenges we identified earlier with hybrid clouds we present the following novel contributions in this paper:
\begin{itemize}
	\item Algorithmic approaches to the data partitioning problem for a hybrid cloud in which we consider the cost of encrypted data storage on a public cloud.
	\item An efficient distributed query optimization and processing engine. Our engine takes into account the cost of querying over encrypted data.
	\item Implementation of these functions as an add-on for a Hadoop and Hive based computing infrastructure.
\end{itemize}

The rest of the paper is organized as follows: Section~\ref{sec:rw} reviews related work in the area of secure distributed data processing. Section~\ref{sec:dp} presents details of our approach to data processing in a hybrid cloud. In section~\ref{sec:exp} we present results of experiments conducted on our implementation. Finally, we present our conclusions and future work in section~\ref{sec:cfw}.
\section{Related Work}
\label{sec:rw}
In this section we provide a brief overview of the relevant research areas that are related to our work in this paper.

A lot of research has focused on data partitioning and distributed query processing without explicitly considering the cost of data security. In our work, we categorically include this cost in both these areas.

\textbf{Data Partitioning:} A lot of research has focused on the problem of data partitioning in single \cite{AgrawalNY04} and distributed systems \cite{RaoZML02} using a strategy such as that given in \cite{GhandeharizadehD90}. Reference \cite{Curino10} uses a graph-based, data-driven partitioning approach for transactional workloads. Our work explicitly considers the cost of querying encrypted attributes that will be stored on the public cloud as a result of the data partitioning process.

\textbf{Distributed Query Processing:} Research efforts have also been made in the area of distributed query processing in the cloud as given in \cite{LogothetisY08}. Distributed query processing has evolved from systems such as SDD-1 \cite{RothnieBFGHLRSW80} that assumed homogeneous databases to DISCO \cite{TomasicRV96} that operated on heterogeneous data sources and finally to Internet scale systems such as Astrolabe \cite{RenesseBV03}. Since we need to execute queries over partitions containing unencrypted and encrypted data, we may not be able to process a query entirely on a public or private cloud. This leads to a cost model that is different from models that currently exist in literature.  

\textbf{Privacy:} The area of privacy-preserving query processing has also received much attention \cite{Curino11, HacigumusILM02}. A homomorphic encryption based technique can be used to query over encrypted data \cite{GennaroGP10} but is expensive when the data size increases. We use techniques given in \cite{HacigumusILM02} to preserve security of data. However, the difference between our work and \cite{HacigumusILM02} is that we can store and query data locally unlike \cite{HacigumusILM02}.

We have also identified a recent work, called Relational Cloud \cite{Curino11}, that attempts to address the problems we have identified above. The difference between our work and Relational Cloud is that our data partitioning scheme considers the cost of querying encrypted attributes stored on a public cloud. Relational Cloud partitions data using a graph-based partitioning scheme without attaching any query cost constraints. These partitions are then encrypted with multiple layers of encryption and stored on a server. A query is then executed over the encrypted data with multiple rounds of communication between a client and server without considering the cost of decrypting intermediate relations. In our work, we explicitly consider the cost of queries that involve all three components of a hybrid cloud: a query over data in a private cloud, a query over non-sensitive (i.e., unencrypted) data and, a query over sensitive (i.e., encrypted) data on a public cloud. To the best of our knowledge, ours is the first work to explicitly estimate the cost of querying over unencrypted and encrypted data in a distributed setting.     
\section{Secure Data Processing}
\label{sec:dp}
This section focuses on the security layer, the query processing engine and the data design layer in that order. This is done since each successive layer is dependent on the concepts presented in the earlier layer.
\subsection{Data Security} 
The \textit{data security} layer prevents the cloud service provider from being able to gain any meaningful information from the sensitive data. The main challenge in this layer is the efficient execution of queries over encrypted data stored on a public cloud. We use the techniques given in \cite{HacigumusILM02} to solve this problem. However, we can store and query data in a private cloud which was not possible in \cite{HacigumusILM02}. We provide a brief overview of certain important concepts from \cite{HacigumusILM02} and refer the reader to \cite{HacigumusILM02} for a more detailed explanation. The query processing engine makes use of these techniques to perform query rewriting that allows an input query to be split into multiple sub-queries over a hybrid cloud.
\subsubsection{Sensitive Attribute Encryption and Storage}
\label{subsec:saes}
The given set of relations, $R_1, R_2, \ldots, R_m$ is vertically partitioned into a partition stored on a user's private cloud and a partition stored on a public cloud. The public cloud partition is further fragmented into a fragment containing sensitive data and a fragment containing non-sensitive data. We now explain how the fragment containing sensitive data is encrypted and stored on a cloud service provider such that queries can be run directly over the encrypted data.

The domain ($\mathcal{D}_i$) of a sensitive attribute $R^{pu}.A_j$ is divided into $z$ partitions, $p_1, p_2, \ldots, p_z$ such that all partitions taken together cover $\mathcal{D}_i$ and no two partitions overlap one another. An identification function, $ident$, assigns an identifier, $ident_{R^{pu}.A_j}(p_z)$ to each partition of $R^{pu}.A_j$ such that $ident_{R^{pu}.A_j}(p_y) \neq ident_{R^{pu}.A_j}(p_z)$ if $y \neq z$. A mapping function, $Map$, is used to map a value $v$ in $\mathcal{D}_i$ of attribute $R^{pu}.A_j$ to the identifier of the partition to which $v$ belongs: $Map_{R^{pu}.A_j}(v) = ident_{R^{pu}.A_j}(p_z)$. A value $v$ of a sensitive attribute $R^{pu}.A_j$ in a vertical partition of a relation, $R^{pu}$ is then encrypted as $E(v) = \langle encrypt(v), Map_{R^{pu}.A_j}(v) \rangle$. We use AES \cite{aes} in CTR mode \cite{ctr} as the encryption function, $E$, while $Map_{R^{pu}.A_j}(v)$ acts as an index on the attribute $R^{pu}.A_j$. The corresponding decryption function, $D$, then decrypts $E(v)$ to return the original value, $t$, after dropping the identifier that is stored along with the encrypted value.
\subsubsection{Mapping query conditions}
When a query is to be evaluated on a public cloud, the query conditions need to be mapped to conditions over the encrypted data stored on this public cloud. For example, in a selection operation with an equality condition, $A_i = v$, the value, $v$, is mapped to the identifier of the partition that contains $v$ as $Map_{A_i}(v)$. Similar mapping conditions exist for other query conditions as shown in \cite{HacigumusILM02}.
\subsubsection{Relational operators over Partitioned Relations}
For a query to execute over partitioned relations that contain sensitive attributes, the underlying relational algebra operators need to be modified to be able to use the functions $ident$ and $Map$. For example if the selection condition contains a sensitive attribute, a partial result can be computed on the public cloud using the index for sensitive attributes, $R_{int} = \sigma_{Map_{c}(C)}^{pu}(R^{pu})$. This result can then be passed back to the query processing engine where it is filtered after being decrypted for tuples that do not match the selection condition, $\sigma_C(R) = \sigma_C(D(R_{int}))$. The remaining relational algebra operators are implemented in \cite{HacigumusILM02}.
\subsection{Query Processing Engine} 
\label{subsec:qpe}
This section describes our query optimization and processing engine. We first present our distributed query execution cost model and then we give a query processing algorithm that is used to query a hybrid cloud. 
\subsubsection{Distributed Cost Model}
\label{subsubsec:dcm}
Algorithm~\ref{alg:qpc} is used to estimate the execution cost of $Q$ queries using statistics for all relations. The execution cost of a query $q_i$ in a hybrid cloud setting can be computed as: 
\begin{equation}\label{eqn:qpc}
c_i = freq(q_i) \times ( max( T^{pu}_{l}, T^{pr}_{l} ) + T_{co} ), 
\end{equation}
where $T^{pu}_l$ and $T^{pr}_l$ are the local processing times on a public and private cloud respectively, and $T_{co}$ is the time to combine the intermediate results at the private cloud. The access frequency of a query $q_i$ is given by $freq(q_i)$. 
\newcommand{\qpc}{\ensuremath{\mbox{\sc QPC}}}
\renewcommand{\algorithmicrequire}{\textbf{Input:}}
\renewcommand{\algorithmicensure}{\textbf{Output:}}
\begin{algorithm}
\caption{$\qpc$()}
\algorithmicrequire \mbox{ $Q$, $SR$}
\algorithmicensure \mbox{ Query execution cost, $c$}
\label{alg:qpc}
\begin{algorithmic}[1]
\STATE{$c \gets 0$}
\FOR{$i \gets 1$ \TO $Q.length$}
	\STATE{$c_i \gets 0$}
	\STATE{Divide $q_i$ into $q_i^{pu}$ and $q_i^{pr}$ using transformation rules}
	\FOR{$j \gets 1$ \TO $q_i^{pu}.length$}
		\IF{$A_k \in R^{pu}$ in $q_j$ is sensitive}
			\STATE{$Map_{A_k}(v) = ident_{A_k}(p_k)$ where $v \in \mathcal{D}_k$ of $A_k$}
		\ENDIF
		\STATE{$c_i \gets c_i + w_2 \times \cardl R^{pu}_{l} \cardr  + w_3 \times \cardl R^{pu}_{tmp} \cardr$} \COMMENT{Compute public cloud cost}
	\ENDFOR
	\FOR{$j \gets 1$ \TO $q_i^{pr}.length$}
		\STATE{$c_i \gets c_i + w_1 \times \cardl R_{tmp}^{pr} \cardr$} \COMMENT{Compute private cloud cost}
	\ENDFOR
	\STATE{$c_i \gets c_i + w_4 \times \cardl R^{pu}_{tmp} + R^{pr}_{tmp} \cardr$} \COMMENT{Combination cost}
	\STATE{$c \gets c + freq(q_i) \text{ } \times c_i$}
\ENDFOR
\RETURN{$c$}
\end{algorithmic}
\end{algorithm}

Early distributed cost models only considered minimizing the communication cost \cite{selinger-adiba-80}. However, we believe that Equation~\ref{eqn:qpc} is a generalized way to estimate the query execution cost since the communication cost has only improved over time \cite{ozsu}. Further, to the best of our knowledge, our cost model is the first to estimate the cost of query execution over unencrypted and encrypted data. The private cloud processing time is estimated as: $T^{pr}_l = w_1 \times \cardl R^{pr}_{tmp} \cardr$ (line 12). $\cardl R^{pr}_{tmp} \cardr$ is the size of the intermediate relation at the private cloud and weight $w_1$ is estimated based on the private cloud infrastructure. The public cloud processing time is evaluated as: $T^{pu}_l = w_2 \times \cardl R^{pu}_{l} \cardr  + w_3 \times \cardl R^{pu}_{tmp} \cardr$ (line 9). $R^{pu}_{l}$ is the size of the relations over which $q_i^{pu}$ will be executed and $\cardl R^{pu}_{tmp} \cardr$ is the size of the intermediate relation created as a result of executing $q_i^{pu}$. Weight $w_2$ is estimated from the public cloud infrastructure while $w_3$ is estimated from the network used to transfer data between the public and private clouds. Finally, $T_{co} = w_4 \times \cardl R^{pu}_{tmp} + R^{pr}_{tmp} \cardr$ represents the time to combine intermediate results at the private cloud (line 14). Weight $w_4$ is estimated based on the private cloud's capability to combine public cloud results with local results. $w_4$ also captures the time to decrypt relations and filter unwanted tuples. $\cardl R_{tmp} \cardr$ is estimated based on the query operator type. Our current work only supports simple SQL queries while we leave the support of nested queries as future work. The running time of this algorithm is $O(k)$ where $k = \sum_{i=1}^{n} q_i^{pu}.length + q_i^{pr}.length$ and $n = Q.length$.
\subsubsection{Distributed Query Processing}
\begin{figure*}[htp]
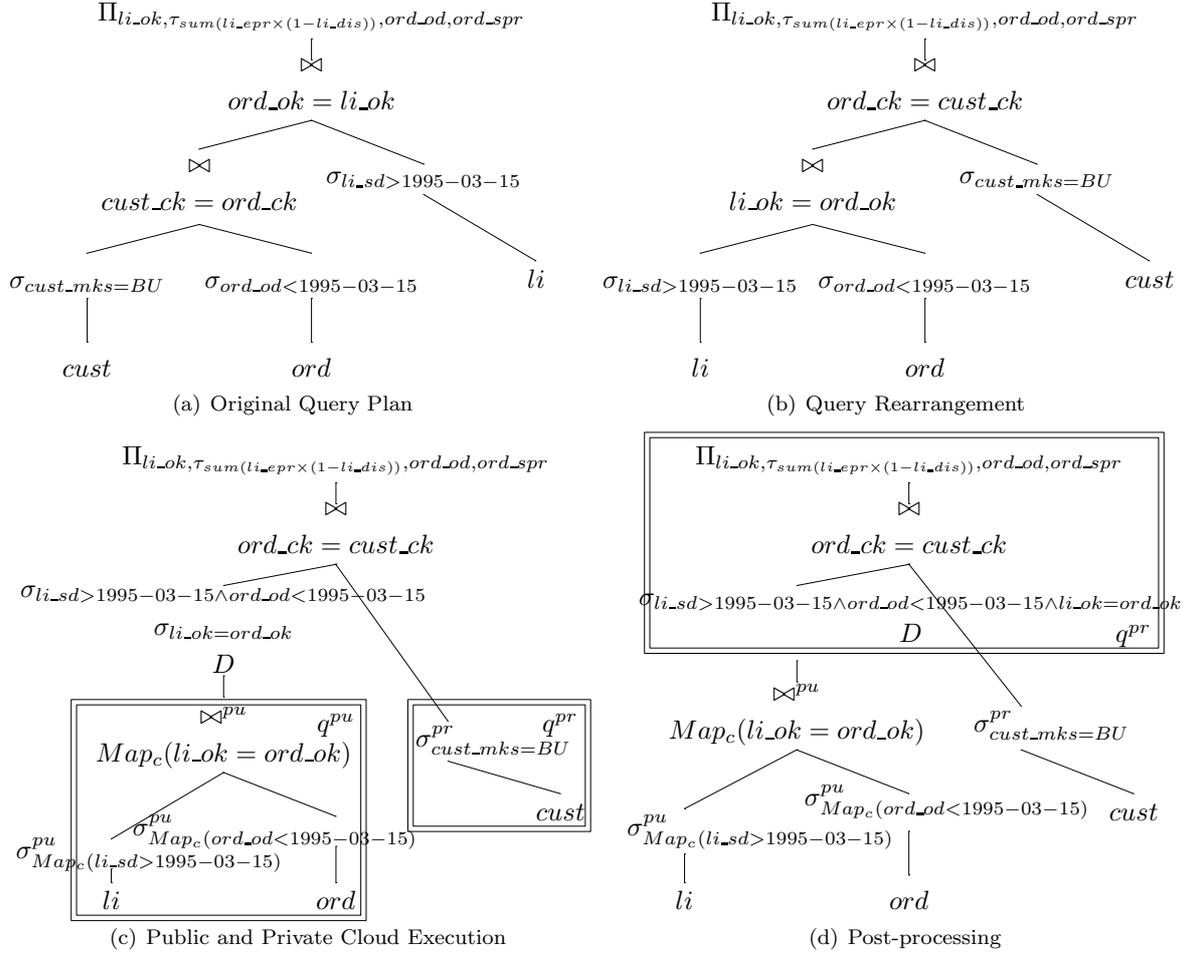

\centering 
\subfigure[Original Query Plan]
{
\label{subfig:exa}
\Tree[1]
{
	& & \K{$\Pi_{li\_ok, \tau_{sum(li\_epr \times (1 - li\_dis))}, ord\_od, ord\_spr}$} \Bk{-1}{2}{d} & & \\
	& & \Kk[0]{5}{$\bowtie$} \Below{$ord\_ok = li\_ok$} \Bk{0}{2}{dl} \DR & & \\
	& \Kk[0]{3.5}{$\bowtie$} \Below{$cust\_ck = ord\_ck$} \Bk{-2}{0}{dl} \Bk{-2}{0}{dr} & & \Kk[0]{2}{$\sigma_{li\_sd > 1995-03-15}$} \Bk{2}{-1}{dr} & \\
	\K{$\sigma_{cust\_mks=BU}$} \D & & \K{$\sigma_{ord\_od < 1995-03-15}$} \D & & \K{$li$} \\
	\K{$cust$} & & \K{$ord$} & &
}
}
\subfigure[Query Rearrangement]
{ 
\label{subfig:exb}
\Tree[1]
{
	& & \K{$\Pi_{li\_ok, \tau_{sum(li\_epr \times (1 - li\_dis))}, ord\_od, ord\_spr}$} \Bk{-1}{2}{d} & & \\
	& & \Kk[0]{5}{$\bowtie$} \Below{$ord\_ck = cust\_ck$} \Bk{0}{2}{dl} \DR & & \\
	& \Kk[0]{3.5}{$\bowtie$} \Below{$li\_ok = ord\_ok$} \Bk{-2}{0}{dl} \Bk{-2}{0}{dr} & & \Kk[0]{2}{$\sigma_{cust\_mks=BU}$} \Bk{2}{-1}{dr} & \\
	\K{$\sigma_{li\_sd > 1995-03-15}$} \D & & \K{$\sigma_{ord\_od < 1995-03-15}$} \D & & \K{$cust$} \\
	\K{$li$} & & \K{$ord$} & &
}
}
\subfigure[Public and Private Cloud Execution]
{
\label{subfig:exc}
\Tree[1]
{
	& & \K{$\Pi_{li\_ok, \tau_{sum(li\_epr \times (1 - li\_dis))}, ord\_od, ord\_spr}$} \Bk{-1}{2}{d} & & \\
	& & \Kk[0]{5}{$\bowtie$} \Below{$ord\_ck = cust\_ck$} \Bk{0}{3}{dl} \Bk{0}{-15}{dr} & & \\
	& \Kk[0]{6}{$\sigma_{li\_sd > 1995-03-15 \wedge ord\_od < 1995-03-15}$} \Below{$\sigma_{li\_ok = ord\_ok}$} \Below{$D$} \Bk{-3}{0}{d} & & & \\
	& \Kk[0]{1}{$\bowtie^{pu}$} \Below{$Map_{c}(li\_ok = ord\_ok)$} \Bk{-4}{-7}{dl} \Bk{-4}{-4}{dr} & \K{$q^{pu}$} & \Kk[6]{-2}{$\sigma^{pr}_{cust\_mks=BU}$} \Bk{-2.5}{-1}{dr} & \K{$q^{pr}$} \\
	\Kk[5]{-5}{$\sigma^{pu}_{Map_{c}(li\_sd > 1995-03-15)}$} \Bk{-5}{-1}{d} & & \Kk[-8]{-2}{$\sigma^{pu}_{Map_{c}(ord\_od < 1995-03-15)}$} \Bk{-2}{-1}{d} & & \K{$cust$} \\
	\K{$li$} & & \K{$ord$} & & \QS[=]{4,1}{6,3} \QS[=]{4,4}{5,5}
}
}
\subfigure[Post-processing]
{ 
\label{subfig:exd}
\Tree[1]
{
	& & \K{$\Pi_{li\_ok, \tau_{sum(li\_epr \times (1 - li\_dis))}, ord\_od, ord\_spr}$} \Bk{-1}{2}{d} & & \\
	& & \Kk[0]{5}{$\bowtie$} \Below{$ord\_ck = cust\_ck$} \Bk{0}{3}{dl} \Bk{0}{-13}{dr} & & \\
	& \Kk[15]{5}{$\sigma_{li\_sd > 1995-03-15 \wedge ord\_od < 1995-03-15 \wedge li\_ok = ord\_ok}$} \Below{$D$} \Bk{-1}{2}{d} & & & \K{$q^{pr}$} \\
	& \Kk[0]{4}{$\bowtie^{pu}$} \Below{$Map_{c}(li\_ok = ord\_ok)$} \Bk{-1}{-3}{dl} \Bk{-1}{-1}{dr} & & \Kk[4]{0}{$\sigma^{pr}_{cust\_mks=BU}$} \Bk{-1}{-1}{dr} & \\
	\Kk[10]{-2}{$\sigma^{pu}_{Map_{c}(li\_sd > 1995-03-15)}$} \Bk{-3}{-1}{d} & & \Kk[5]{2}{$\sigma^{pu}_{Map_{c}(ord\_od < 1995-03-15)}$} \D & & \K{$cust$} \\
	\K{$li$} & & \K{$ord$} & & \QS[=]{1,1}{3,5}
}
}
\caption{Query rewriting for a join query}
\label{fig:ex}
\end{figure*}

Algorithm~\ref{alg:qpe} presents details of our query processing engine. This algorithm is used to execute a query over a hybrid cloud. Algorithm~\ref{alg:qpe} consists of four phases, each of which we motivate with the query given in Figure~\ref{fig:ex} that is a modified version of Q3 of TPC-H \cite{tpc-h} and is given as follows:
\newcommand{\qpe}{\ensuremath{\mbox{\sc QPE}}}
\renewcommand{\algorithmicrequire}{\textbf{Input:}}
\renewcommand{\algorithmicensure}{\textbf{Output:}}
\begin{algorithm}
\caption{$\qpe$()}
\algorithmicrequire \mbox{ $q_i$}
\algorithmicensure \mbox{ Query result, $R_{res}$}
\label{alg:qpe}
\begin{algorithmic}[1]
\STATE{Divide $q_i$ into $q_i^{pu}$ and $q_i^{pr}$ using transformation rules}
\STATE{Execute $q_i^{pu}$ and $q_i^{pr}$ in parallel}
\FOR{$j \gets 1$ \TO $q_i^{pu}.length$}
	\IF{$A_k \in R^{pu}$ in $q_j$ is sensitive}
		\STATE{$Map_{A_k}(v) = ident_{A_k}(p_k)$ where $v \in \mathcal{D}_k$ of $A_k$}
	\ENDIF
	\STATE{$R_{tmp}^{pu} \gets $ Execute $q_j$ over $R^{pu}$} \COMMENT{Public cloud execution}
\ENDFOR
\FOR{$j \gets 1$ \TO $q_i^{pr}.length$}
	\STATE{$R_{tmp}^{pr} \gets $ Execute $q_j$ on $R^{pr}$} \COMMENT{Private cloud execution}
\ENDFOR
\STATE{$R_{res} \gets $ Combine $R_{tmp}^{pu}$ and $R_{tmp}^{pr}$} \COMMENT{Result combination}
\RETURN{$R_{res}$}
\end{algorithmic}
\end{algorithm}

\small
\begin{verbatim}
SELECT li_ok, sum(li_epr*(1-li_dis)), ord_od, ord_spr
FROM cust, ord, li
WHERE cust_mks=BU AND cust_ck = ord_ck AND li_ok = ord_ok AND ord_od < 1995-03-15 AND li_sd > 1995-03-15
\end{verbatim}
\normalsize
We assume that the lineitem and order relations are sensitive and hence are encrypted on a public cloud while the customer relation is stored on a private cloud. Figure~\ref{subfig:exa} shows an execution plan for the query which is transformed using the different phases of Algorithm~\ref{alg:qpe} as given below: \\
\textbf{Query Rearrangement}: To divide a query $q$ into $q^{pu}$ and $q^{pr}$ we use relational algebra transformation rules. This phase will transform the query plan from Figure~\ref{subfig:exa} to the plan given in Figure~\ref{subfig:exb} using the commutativity rule of the join operation for our example. \\
\textbf{Public Cloud Execution}: In this phase a generated sub-query(ies) is(are) executed over relations in a public cloud. Figure~\ref{subfig:exc} shows how the query plan is divided into a public cloud query, $q^{pu}$, and a private cloud query, $q^{pr}$. We push as much processing to a public cloud as possible by mapping the original query conditions to conditions over encrypted attributes using the $Map$ function. The box on the left hand side of Figure~\ref{subfig:exc} represents $q^{pu}$ for our example. \\
\textbf{Private Cloud Execution}: A sub-query(ies) is(are) directly executed over relations in a private cloud. The box on the right in Figure~\ref{subfig:exc} represents the private cloud query, $q^{pr}$, for our example. This query is executed in parallel with $q^{pu}$ on the public cloud. \\
\textbf{Post-processing}: This phase combines the intermediate relations generated at the public and private clouds into the final result. Figure~\ref{subfig:exd} shows the post-processing step ($q^{pr}$) for our example query. This step decrypts the data that is received from the public cloud. Next, incorrect results are filtered from the decrypted data by applying the original query conditions. Finally, the results from the public and private clouds are combined and returned to a user. 
\subsection{Data Design Layer} 
The \textit{data design} layer is concerned with partitioning a set of relations between a user's private cloud and a public cloud service provider. The process of partitioning is necessary since a user's private cloud may not have sufficient storage and/or processing power. This process becomes more complex in our setting since we want to protect the privacy of a user's data. We first define the data design problem and then present algorithmic strategies to solve this problem.
\subsubsection{Data Partitioning Problem}
The data design problem in a hybrid cloud setting is defined as follows: Minimize the execution cost of a set of queries, $Q$, over a distribution of attributes, $A$, among the public ($A^{pu}$) and private ($A^{pr}$) clouds. This problem is subject to the condition that $A^{pr}.size \le W$, where $W$ is the disk space available on the private cloud. This is clearly an optimization problem, which we call the CLOUD-SUBSET-SELECTION (CSS) problem. There are an exponential number of subsets of $A$, each of which needs to be tested as a solution to the CSS problem. We can verify in polynomial time that the execution cost of $Q$ queries is less than a bound $C$ for a given $A^{pu}$ and $A^{pr}$. Hence, the CSS problem belongs to the class of NP problems. Moreover, the 0-1 Knapsack Problem (0-1 KP) can be reduced to the CSS problem making the CSS problem NP-complete. A formal proof of NP-completeness is given in Appendix~\ref{app:npc}.
\subsubsection{Algorithmic Solutions to Data Partitioning}
\label{subsubsec:algsoln}
We use two different algorithmic strategies to produce a close to optimal solution for the CSS problem based on dynamic programming and hill climbing. The idea of using a dynamic programming solution to solve the CSS problem comes from the similarity between the CSS problem and 0-1 KP. Further, the hill climbing technique is inspired from the SDD-1 algorithm \cite{RothnieBFGHLRSW80}, which is a well known query optimization algorithm for distributed databases.
\newcommand{\cssdp}{\ensuremath{\mbox{\sc CSS-DP}}}
\renewcommand{\algorithmicrequire}{\textbf{Input:}}
\renewcommand{\algorithmicensure}{\textbf{Output:}}
\begin{algorithm}
\caption{$\cssdp$()}
\algorithmicrequire \mbox{ $A$, $W$, $Q$, $SR$}
\algorithmicensure \mbox{ An array $p$}
\label{alg:cssdp}
\begin{algorithmic}[1]
\STATE{Update $SR$ such that $A^{pu} = A$, $A^{pr} = \emptyset$}
\STATE{$c_{in} \gets $ QPC($Q$, $SR$)} \COMMENT{Compute initial cost}
\FOR{$i = 0$ \TO $W$}
\STATE{$p[0,i] \gets 0$}
\ENDFOR
\FOR{$i \gets 1$ \TO $A.length$}
	\STATE{$p[i,0] \gets 0$}
	\FOR{$j = 1$ \TO $W$}
		\IF{$A_i.size \le j$}
			\STATE{Update $SR$ such that $A_i$ moved from $A^{pu}$ to $A^{pr}$}
			\STATE{$c_i \gets c_{in} - $ QPC($Q$, $SR$)} \COMMENT{Compute the profit for $A_i$ such that $A^{pu} = A^{pu} - A_i$, $A^{pr} = A_i$}
			\IF{$c_i + p[i-1,j-A_i.size] > p[i-1,j]$}
				\STATE{$p[i,j] \gets p_i + p[i-1,j-A_i.size]$ }
			\ELSE
				\STATE{$p[i,j] \gets p[i-1,j]$}
			\ENDIF
		\ELSE 
			\STATE{$p[i,j] \gets p[i-1,j]$}
		\ENDIF
	\ENDFOR
\ENDFOR
\RETURN{$p$}
\end{algorithmic}
\end{algorithm}

Algorithm~\ref{alg:cssdp} (CSS-DP) is derived from a dynamic programming solution to 0-1 KP. However, there are several differences between the two problems. Firstly, 0-1 KP considers items with a weight $w_i$ and a profit $p_i$, while the CSS problem considers attributes with a size $A_i.size$ and an associated cost $c_i$. 0-1 KP tries to maximize the profit $P$ while maintaining the weight of the knapsack less than $W$. The CSS problem tries to minimize the execution cost of $Q$ queries given that $A_i$ is placed in the private cloud under the constraint that $A^{pr}.size \le W$. The algorithm takes the following parameters as input: a set of attributes $A$, the size of the private cloud $W$, a set of input queries $Q$ and the statistics for all relations as a set, $\forall R_i \in R \text{; } stat(R_i) \in SR$. Note that $SR$ captures the partitioning of $A$ into $A^{pu}$ and $A^{pr}$. Algorithm~\ref{alg:cssdp} begins by calling Algorithm~\ref{alg:qpc} to compute an initial cost ($c_{in}$) for $Q$ queries given that $A^{pu} = A$ and $A^{pr} = \emptyset$ (line 2). Then, the profit associated with each $A_i$ can be computed as: $c_i = c_{in} -$ QPC($Q, SR$) using an updated $SR$ such that $A_i$ is moved from $A^{pu}$ to $A^{pr}$ (line 11). Algorithm~\ref{alg:cssdp} then finds the maximum profit that can be achieved over $A$ using the profit for each $A_i \in A$. The term ``profit'' is used under the assumption that the private cloud is able to process queries faster than the public cloud. However, if the converse is true, there could be a loss in execution cost. Then, the algorithm may keep most of the attributes in the public cloud. After execution, Algorithm~\ref{alg:cssdp} returns an array of size $n \times W$, where $n = A.length$, that contains the maximum profit that can be achieved in position $[n,W]$. The set $A^{pr}$ can be computed by starting at $p[n,W]$ and tracing backwards based on the profit earned and size associated with every attribute. Note that the running time of this algorithm $O(nW \times k)$, where $k$ is the running time of Algorithm~\ref{alg:qpc}, and the time to compute the set $A^{pr}$ is $O(n)$.

\newcommand{\csshc}{\ensuremath{\mbox{\sc CSS-HC}}}
\renewcommand{\algorithmicrequire}{\textbf{Input:}}
\renewcommand{\algorithmicensure}{\textbf{Output:}}
\begin{algorithm}
\caption{$\csshc$()}
\algorithmicrequire \mbox{ $A$, $W$, $Q$, $SR$, $bound$}
\algorithmicensure \mbox{ $A^{pu}$}
\label{alg:csshc}
\begin{algorithmic}[1]
\STATE{$A^{pr} \gets A$}
\STATE{Create $A^{pu}$ using selected strategy.} \COMMENT{Initial solution}
\STATE{Update $SR$ based on $A^{pu}$ and $A^{pr}$}
\STATE{$c_{in} \gets$ QPC($Q$, $SR$)} \COMMENT{Compute initial cost}
\STATE{$c_{prev} \gets c_{in}$; $c_{new} \gets c_{in} + 1$}
\WHILE{$c_{new} > c_{prev} \text{ } || \text{ } iter \le bound$}
	\STATE{$iter \gets iter + 1; \text{ } c_{prev} \gets c_{new}$}
	\STATE{Randomly swap a pair of attributes from $A^{pu}$ and $A^{pr}$ to get $A_{n}^{pu}$ and $A_{n}^{pr}$ such that $A_n^{pr} \le W$}
	\STATE{Update $SR$ based on $A_{n}^{pu}$ and $A_{n}^{pr}$}
	\STATE{$c_{new} \gets$ QPC($Q$, $SR$)} \COMMENT{Compute new cost}
\ENDWHILE
\STATE{$A^{pu} \gets A_{n}^{pu}$}
\RETURN{$A^{pu}$}
\end{algorithmic}
\end{algorithm}
Algorithm~\ref{alg:csshc} (CSS-HC) uses a hill climbing technique and takes the same input parameters as CSS-DP. It also takes a $bound$ on the number of random swaps to perform between $A^{pu}$ and $A^{pr}$. An initial solution is built (line 2) using one of the following greedy strategies: 1) Keep as many attributes from the query set in the private cloud as possible (CSS-HC-Query). 2) Keep as many sensitive attributes in the private cloud as possible (CSS-HC-sensitivity). An initial execution cost ($c_{in}$) of $Q$ queries is then computed using Algorithm~\ref{alg:qpc} (line 4). A pair of attributes from $A^{pu}$ and $A^{pr}$ is then randomly swapped (line 8) and the execution cost is recomputed (line 10). If this cost is better than the previous cost the new partitioning is retained. If the converse is true, the process of swapping attributes and recomputing execution cost is repeated. The total running time of the algorithm is $O(bound \times k)$ where $k$ represents the running time of Algorithm~\ref{alg:qpc}.
\section{Experimental Results}
\label{sec:exp}
This section presents the results of experiments conducted to compare the performance of the previous two algorithms. We first present details of our experimental setup followed by the set of experiments.

\textbf{Experimental Setup}: Our experiments were conducted on two local clusters that are on different sub-networks of the same university intranet. We consider that this configuration simulates a real-world hybrid cloud well. This is because the average transfer speed between the nodes of our two local clusters ($\approx$ 672.04KB/sec) is the same as the average transfer speed between a node in our local network with Amazon S3\footnote{http://aws.amazon.com/s3/} ($\approx$ 683.05KB/sec for encrypted data and $\approx$ 710.46KB/sec for unencrypted data) \cite{rest}. The first cluster consists of 4 nodes each with a Pentium IV processor with a 250 GB hard drive and 1GB of main memory and is used as the private cloud. The second cluster consists of 20 nodes each with a Pentium IV processor with 290GB to 360GB disk space and 4GB main memory and is used as the public cloud. Both clusters are setup using Hadoop\footnote{http://hadoop.apache.org/} v0.20.2 and Hive\footnote{http://hive.apache.org/} v0.6.0. The first cluster is configured with $\approx$ 350GB disk space while the second is configured with $\approx$ 4.7TB of disk space for HDFS.

\textbf{Security Functions}: We used SHA-256 \cite{sha-256} as the $ident$ function. Further, we used the built-in datatypes, int, double and string in Hive to represent the attributes of all relations from the TPC-H benchmark. The number of partitions, $P$, for a datatype could be varied from 1 to the number of unique values in the domain of an attribute. When $P = 1$ we get a high degree of security, however, query processing time increases since all values are mapped to the same partition. When $P = $ the number of unique values in the domain of an attribute, query processing is fast since a small subset of values is mapped to a partition. However, the level of security is reduced since for example, a public cloud service provider could learn the data access patterns of queries. We used the following equation to determine $P$:
\begin{equation}
P = \frac{log(max - min)}{log \text{ } 2},
\end{equation}
where $min$ and $max$ represent the minimum and maximum values for the datatype as mandated by the TPC-H benchmark. For an integer datatype, $min = -2,147,483,646$ and $max = 2,147,483,647$. This leads to $n = 31$ partitions using the above equation. Similarly, for a double datatype, $min = -9,999,999,999.99$ and $max = 9,999,999,999.99$, leading to $n = 34$ partitions. For the string datatype, we created 36 partitions as $a-z$ and $0-9$. Unless specified otherwise, we use these partitions in our experiments. To encrypt subsets of attributes we used the AES \cite{aes} in CTR mode \cite{ctr} from the Java cryptographic extension.

\textbf{Queries}: We have used the TPC-H benchmark \cite{tpc-h} with a scale factor 300 ($\approx$ 323GB) in our experiments. We did not run experiments for larger databases since we think the current case gives us sufficient insight into the workings of the algorithmic strategies. The first experiment used Q10 of TPC-H without the grouping and aggregate operations. The next two experiments used a query workload of 100 queries containing modified versions of TPC-H queries Q1, Q3, Q6 and Q10. In particular, we do not perform grouping and aggregate operations in any query. Further, $freq(q_i)$ was randomly selected between 1 and 1000. Additional details of the workload preparation are given in Appendix~\ref{app:qwp}.

\textbf{Preliminary Experiments:} We ran a set of preliminary experiments to estimate the weights defined in our cost model. These experiments were run only once and generate weights that are effective as will be shown. The values generated were: $w_1 = 0.000545146$, $w_2 = 0.000072686$, $w_3 = 0.000001488$ and $w_4 = 0.0000041$. Details of these preliminary experiments are provided in Appendix~\ref{app:pe}.
%
\begin{figure}[htp]
\centering
\includegraphics[width=80mm,height=60mm]{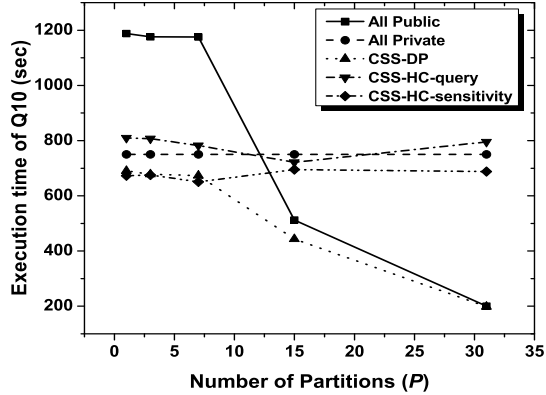}
\hfill
\caption{Comparison of CSS-DP and CSS-HC when the number of partitions, $P$, is varied}
\label{fig:changeP}
\end{figure} 

\textbf{Experiment with changing the number of partitions ($P$):}  The aim of this experiment was to compare the algorithms when $P$ is increased while the private cloud size ($\approx$ 90GB) and the sensitivity of attributes ($\approx$ 50\%) are fixed\footnote{Our running times are very similar to the timings in \cite{hive-tpc-h} even though we use a less powerful cluster than \cite{hive-tpc-h}.}. Further, we wanted to show that although $w_1 > w_2$, the choice of $P$ affects the query performance on the public cloud. The solid and dashed lines in Figure~\ref{fig:changeP} represent the running times when $A^{pu} = A$ (All-Public) and $A^{pr} = A$ (All-Private) respectively. The running time is constant for All-Private as it does not use partitions for query processing. For All-Public, the time decreases as $P$ is increased. When $P$ is small, the query takes longer to perform a join on the public cloud, since a large number of values map to the same partition. More time is also spent in transferring data to, and decrypting data on, the private cloud. As $P$ increases, the time needed to perform a join as well as the transfer and decryption time reduces. Both the CSS-HC techniques perform similar to All-Private. This is because they always leave attributes from the query set or sensitive attributes, in the private cloud. When $P$ is small, CSS-DP picks attributes such that $A_j \in A^{pr}$ when $A_j \in Q$. The distribution of data for CSS-DP in such a case, for example for $P = 4$ is: about $76\%$ data in the public cloud and $24\%$ data in the private cloud. This leads to a running time that is better than All-Public and All-Private. However, as $P$ increases more attributes from the query set are pushed to the public cloud. This is because the time taken to perform a join on the public cloud followed by decrypting and filtering intermediate results is much lesser than performing the query on the private cloud.
%
\begin{figure}[htp]
\centering
\includegraphics[width=80mm,height=60mm]{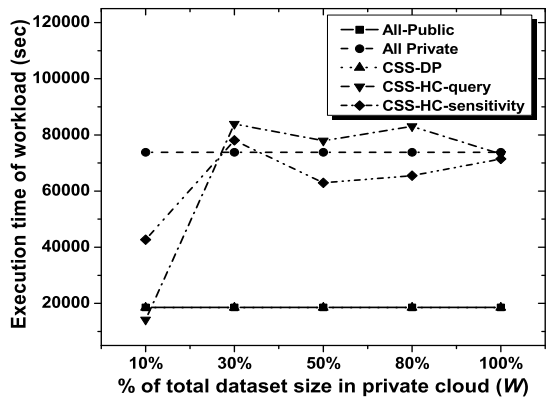}
\hfill
\caption{Comparison of CSS-DP and CSS-HC when the private cloud size, $W$, is scaled}
\label{fig:changeW}
\end{figure} 

\textbf{Experiment with changing the private cloud size ($W$):} The goal of this experiment is to compare the performance of our algorithms when $W$ increases while the sensitivity of attributes is fixed randomly (at $\approx$ 40\% over all relations) and the default partitions of 31, 34 and 36 are used. The dashed line in Figure~\ref{fig:changeW} represents the running time for All-Private. Further, All-Public and CSS-DP overlap in Figure~\ref{fig:changeW}. This is expected since from Figure~\ref{fig:changeP} we see that the running times for these two cases are similar for $P \ge 15$. CSS-DP performs much better than CSS-HC. When an attribute $A_i \in Q$ is moved from $A^{pu}$ to $A^{pr}$, it would result in a loss in execution cost. Therefore, CSS-DP picks attributes such that $A_j \in A^{pr}$ when $A_j \notin Q$. This is expected, since $w_1 > w_2 $, however, this is not a general trend as was shown by the previous experiment. On the other hand, both CSS-HC techniques start with an initial solution that is iteratively improved. However, since $w_1 > w_2$, the initial estimate of execution cost is already much higher than the CSS-DP case. Hence, a random swap of any $A_i \in A$ between $A^{pu}$ and $A^{pr}$ does not change the execution cost significantly. For the $W = 10\%$ case, the CSS-HC techniques perform better than for the other $W$ cases. The CSS-HC techniques store most of the attributes needed by the query set on the public cloud and hence query processing is much faster. As $W$ scales, more attributes from the query set are brought into the private cloud. Therefore, the processing time becomes as slow as the case when $A^{pr} = A$.
%
\begin{figure}[htp]
\centering
\includegraphics[width=80mm,height=60mm]{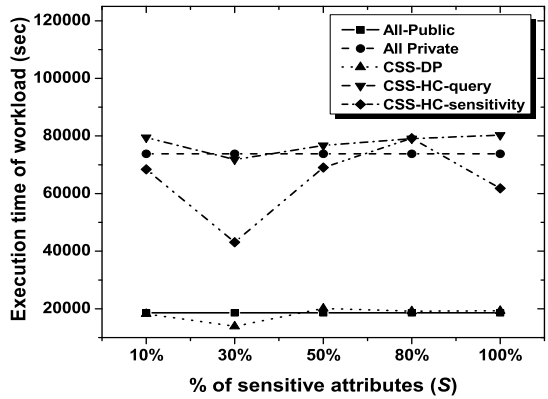}
\hfill
\caption{Comparison of CSS-DP and CSS-HC when attribute sensitivities, $S$, are increased}
\label{fig:changeS}
\end{figure} 

\textbf{Experiment with changing sensitivities ($S$):} This experiment measures the time to run the workload when $W$ is fixed (at $\approx$ 150GB) while $S$ is varied from some attributes being sensitive to all being sensitive. Figure~\ref{fig:changeS} shows a comparison of the algorithmic strategies and we see results that are similar to Figure~\ref{fig:changeW}. CSS-DP selects attributes such that $A_j \in A^{pr}$ when $A_j \notin Q$; this makes the query execution time for the workload much faster. The CSS-HC strategies select an initial solution based on the selected criteria (query or sensitivity). Again, since $w_1 > w_2$, neither of these strategies produces a good workload execution time when compared with CSS-DP. For the $S = 30\%$ case the CSS-HC-sensitivity technique produces a query workload time that is better than the other $S$ cases. This is because the initial solution and subsequent random swaps leave most of the query set attributes in the public cloud.

\textbf{General Observations:} We observe that for some cases of CSS-HC the execution time is higher than when $A^{pr} = A$. This is due to the additional time for the post-processing step. We also observe that the weights estimated by the preliminary experiments perform well. The weights capture the purpose for which they were measured as shown by the weights for private ($w_1$) and public ($w_2$) cloud processing.
\section{Conclusions and Future Work}
\label{sec:cfw}
With the advent of cloud computing, a hybrid cloud may be suitable for users who wish to balance data security with scalable data processing. We have identified three challenges that must be overcome before this approach can be adopted. 

The first challenge deals with data partitioning between a private cloud and a service provider when there are sensitive attributes in the data. We have developed two algorithmic approaches that produce a close to optimal solution to this optimization problem. The second challenge is how to store a user's data securely on a cloud service provider? We have used existing encryption techniques to store a user's sensitive data on the cloud service provider. Moreover, this technique allows us to push most of the query processing work to the cloud service provider without the need of decrypting the stored data. Finally, the last challenge addresses the problem of distributed query processing over unencrypted and encrypted data. We have developed a cost model that estimates the cost of query execution over unencrypted and encrypted data. We have also presented a query processing engine that splits a user query into a public and private cloud query(ies). Each of these query(ies) is(are) then executed at each site using the best available local query plan. 

We are exploring the following areas for future research: 1) We have only considered a vertical partitioning of relations in this paper which will be extended to include horizontal and hybrid partitioning schemes. 2) Our cost model considers only simple SQL queries. We plan to build a more sophisticated model with support for nested queries. 3) In this paper we used Hadoop and Hive as the underlying cloud computing technologies. We aim to extend this work with more experiments into a generalized tool that will work with other existing public cloud services.
\bibliographystyle{unsrt}
\bibliography{CoRR-2011}
\clearpage
\normalsize
\appendix
\section{The CSS Problem is NP-complete}
\label{app:npc}
\noindent \textit{Given}: A set of relations, $R = \{R_{1}, R_{2}, \ldots, R_{m}\}$, a set of attributes, $A = \{A_{1}, A_{2}, \ldots, A_{n}\}$ over all relations $R$ where an attribute $A_{j} \in \{sensitive, non\mbox{-}sensitive\}$, a set of input queries, $Q = \{Q_{1}, Q_{2}, \ldots, Q_{k}\}$. \\

\noindent \textit{Problem}: We have the following optimization problem for CLOUD-SUBSET-SELECTION (CSS),
\begin{equation*}
	  \begin{aligned}
	  & \underset{X}{\text{minimize}} & & \sum_{i=1}^{k} freq(q_i) \times QPC_{q_i}(X) \\
	  & \text{subject to} & & \sum_{j=1}^{n} s(A_j) \times x_j \le \text{ PRIVATE\_CLOUD\_SIZE} \\
	  & \text{where} & & x_j = \left\{ \begin{array}{l l} 1 & \quad \text{if $A_j$ is in the private cloud;}\\ 0 & \quad \text{if $A_j$ is in the public cloud,}\\ \end{array} \right. \\ 
	  & \text{and} & & s(A_j) \text{ denotes the size of attribute $A_j$.} 
	  \end{aligned}
\end{equation*}

\noindent We convert the optimization problem to a decision problem: \\

\noindent \textit{Problem}: Is there a partitioning of $A$ into $A^{pu}$ and $A^{pr}$ such that the cost of executing $Q$ queries over $X = \{A^{pu},A^{pr}\}$ is at most $C$, where $A_j \in A^{pu}$ if the corresponding $x_j = 0$ and $A_j \in A^{pr}$ if the corresponding $x_j = 1$?

\begin{lemma}
\label{ss-np}
The CSS problem belongs to the class NP.
\end{lemma}
\begin{proof}
We shall provide a two-input, polynomial-time algorithm $Al$ that can verify CSS. One of the inputs to the algorithm $Al$ is a set of queries $Q$ while the other input is a certificate corresponding to a partitioning of the attribute set $A$ into $A^{pu}$ and $A^{pr}$.

Algorithm $Al$ is constructed as follows: For each query $q_{i} \in Q$, $Al$ determines the cost of executing $q_{i}$ given the partitions $A^{pu}$ and $A^{pr}$, i.e., $QPC_{q_i}(X)$. We assume that $QPC_{q_i}(X)$ can be computed in polynomial time. If the sum of execution costs of all queries $Q$ is less than the bound $C$, the algorithm outputs 1, since the partitioning of $A$ into $A^{pu}$ and $A^{pr}$ provides a cost less than or equal to the bound C. Otherwise, $Al$ outputs 0.

Whenever a partitioning of $A$ into $A^{pu}$ and $A^{pr}$ that produces an execution cost over all queries $Q$ that is less than or equal to $C$ is input to algorithm $Al$, there is a certificate whose length is polynomial in the size of $A$ and that causes $Al$ to output a 1. Whenever a partitioning of $A$ into $A^{pu}$ and $A^{pr}$ that produces an execution cost greater than $C$ is input, algorithm $Al$ outputs a 0. Algorithm $Al$ runs in polynomial time. Thus, CSS can be verified in polynomial time, and CSS $\in$ NP.
\end{proof}

\begin{lemma}
\label{ss-np-reduction}
The CSS problem is NP-hard.
\end{lemma}
\begin{proof}
To prove that CSS is NP-hard we show that the 0-1 Knapsack Problem (KP) $\le_{P}$ CSS. We show that any instance of 0-1 KP can be reduced in polynomial time to an instance of the CSS problem. \\

\noindent We first define 0-1 KP as follows: Given a set of $n$ items and a $knapsack$, with $p_{j} = profit$ of item $j$, $w_{j} = weight$ of item $j$, $c = capacity$ of the $knapsack$, select a subset of items so as to
\begin{equation*}
	  \begin{aligned}
	  & \text{maximize} & & z = \sum_{j=1}^{n} p_jx_j \\
	  & \text{subject to} & & \sum_{j=1}^{n} w_jx_j \le c, \\
	  & \text{where} & & x_j = \left\{ \begin{array}{l l} 1 & \quad \text{if item $j$ is selected;}\\ 0 & \quad \text{otherwise}\\ \end{array} \right. \\ 
	  & \text{and} & & j \in N = \{1, 2, \ldots, n\}. 
	  \end{aligned}
\end{equation*}
\noindent We convert this problem into the following minimization problem subject to the same conditions as before,
$$\text{minimize } z = \sum_{j=1}^{n} -p_jx_j$$

\noindent 0-1 KP can then be recast as the following decision problem: Can we achieve a profit of at most $P$ without exceeding the weight $c$? \\

\noindent The reduction algorithm begins with an instance of 0-1 KP. Let $X = \{x_1, x_2, \ldots, x_n\}$ be the set of items each of which is associated with a profit $p_j$ and weight $w_j$, where $j \in N = \{1, 2, \ldots, n\}$. Also, let $c$ be the capacity of the \textit{knapsack}. We will construct an instance of the CSS problem with a set of attributes $A$ over all relations $R$ from the set of $n$ items such that the 0-1 KP instance is satisfiable if and only if the CSS instance is satisfiable. Satisfiability in these problems means that the decision problem is answered with a `yes'. The instance of the CSS problem is constructed as follows: \\

For every item $x_j \in X$, the instance of the CSS problem has an attribute $A_j$. Further, the weight $w_j$ of an item $x_j$ corresponds to the size of the attribute $A_j$, i.e., $w_j = s(A_j)$. An initial cost, $c_{in}$ is computed such that $A^{pu} = A$ and $A^{pr} = \emptyset$. Then, the profit of item $x_j$ corresponds to the execution cost of $Q$ queries (denoted as $c_j$) over a partitioning of $A$ into $X$ as given in the minimization problem of CSS such that $A_j \in A^{pr}$. The profit can be computed as: $c_j = c_{in} - \sum_{i=1}^{k} freq(q_i) \times QPC_{q_i}(X)$. This means that $p_jx_j = c_jx_j$ where $x_j = 1$ for $A_j$ in the CSS problem. Then, the total profit $P$ becomes the execution cost, $C$, over all queries $Q$. Also the total capacity of the knapsack, $c$, becomes the size of the private cloud, PRIVATE\_CLOUD\_SIZE, which is computed as $\sum_{j=1}^{n} s(A_j) \times x_j \text{; } \forall x_j = 1$. This instance of CSS can easily be computed from the instance of 0-1 KP in polynomial time. \\

\noindent We now show that this transformation is a reduction under the assumption that $QPC_{q_i}(X)$ can be computed in polynomial time for any query $q_i \in Q$. First, suppose that the given instance of 0-1 KP is satisfiable. Then, we have a subset $\bar{X} \subseteq X$ such that the total profit $\bar{P} = \sum_{j=1}^{n} p_j \forall x \in \bar{X} \le P$. We claim that $\bar{X}$ corresponds to $A^{pr}$. Any element $A_j$ will only be added to $A^{pr}$ when both conditions: $c_j \le C$ and $\sum_{j=1}^{n} s(A_j) \times x_j \le \text{ PRIVATE\_CLOUD\_SIZE}$, hold.

Conversely, suppose that the CSS problem instance has a partitioning of $A$ into $A^{pu}$ and $A^{pr}$ that satisfies all constraints. Every $x_j$ corresponding to an $A_j \in A^{pr}$ can be selected from the instance of 0-1 KP to form the set of elements that will achieve at most profit $P$. This is because, each $x_j$ will achieve at most profit $p_j \le P$ and weight $w_j \le c$.
\end{proof}

\begin{theorem}
\label{ss-np-complete}
The CSS problem is NP-complete.
\end{theorem}
\begin{proof}
Immediate from Lemmas~\ref{ss-np} and~\ref{ss-np-reduction} and the definition of NP. 
\end{proof}
\section{Additional Details of the Experiments}
The TPC-H benchmark is a decision support benchmark that consists of a schema that is typical of any business organization \cite{tpc-h}. The TPC-H benchmark provides a system that inspects large amounts of data by executing queries with a high degree of complexity that are derived from critical business questions. The TPC-H benchmark consists of 8 relations and 22 queries having a realistic context that capture the business activities of a wholesale supplier \cite{tpc-h}.
\subsection{Preliminary Experiments}
\label{app:pe}
A set of preliminary experiments was run to estimate the weights $w_1$ to $w_4$ that are used in out cost model. Weights $w_1$ and $w_2$ represent the local processing times on the private and public clouds respectively and are estimated by running the same set of 4 queries on both the clouds. These queries consist of Q1, Q5 and Q13 from TPC-H \cite{tpc-h} while the $\text{4}^{th}$ query is as follows: 
\begin{verbatim}
select * from lineitem l join orders o on l.l_orderkey = o.o_orderkey
\end{verbatim}
The four queries were selected to have a mix of low and high selectivity. From the average running times of each query we determined $w_1$ as $w_1 = \frac{t_1}{b_1} + \frac{t_2}{b_2} + \frac{t_3}{b_3} + \frac{t_4}{b_4}$. $t_1$ represents the time to run Q1 on the private cloud and $b_1$ represents the number of bytes generated in the result of an execution of Q1. The weight $w_2$ is computed in the same way as $w_1$. Using this procedure we have estimated, $w_1 = 0.000545146$ and $w_2 = 0.000072686$. Since $w_1 > w_2$, the private cloud processing is slower than public cloud processing for our hybrid cloud. However, this is not a general rule and the converse may also be true. Our cost model and partitioning algorithms capture either of these behaviors.

Weight $w_3$ denotes the time required to transfer $R^{pu}_{tmp}$ to the private cloud and is estimated using the following query:
\begin{verbatim}
select * from lineitem limit x,
\end{verbatim}
where we vary x from 10\% to 100\% of the number of tuples in the lineitem relation. This query was selected since the lineitem relation is the largest of all the TPC-H relations and $w_3$ can be best estimated when a large amount of data is transferred. We then estimated $w_3 = 0.000001488$ by averaging the running time of the previous 10 queries.

The weight $w_4$ is used to capture the time taken at the private cloud to combine the intermediate results, $R^{pu}_{tmp}$ and $R^{pr}_{tmp}$, obtained from the public and private clouds respectively. The same query used to estimate $w_3$ was also used to estimate $w_4$. However, the lineitem relation was partitioned between the public and private clouds in the following way:
\begin{verbatim}
Private cloud: l_orderkey l_partkey l_quantity l_linestatus l_shipdate l_shipinstruct
Public cloud sensitive: l_suppkey l_linenumber l_extendedprice l_commitdate l_shipmode
Public cloud non-sensitive: l_disocunt l_tax l_returnflag l_receiptdate l_comment
\end{verbatim}
The weight $w_4 = 0.0000041$ was then estimated by averaging the time to combine the results obtained from the different partitions for the given 10 queries.
\subsection{Query Workload Preparation}
\label{app:qwp}
We prepared a TPC-H database of scale factor 300 ($\approx$ 323GB) using the TPC-H dbgen tool. We also created a query workload of 100 queries using 4 TPC-H queries (Q1, Q3, Q6 and Q10). For each of these queries the predicates in the query are randomly modified to vary the range of the data that is accessed. We summarize the ranges used for each predicate below:
\begin{verbatim}
  1992-01-01 <= l_shipdate <= 1998-12-31
  c_mktsegment = {AUTOMOBILE, BUILDING, FURNITURE, MACHINERY, HOUSEHOLD}
  1992-01-01 <= o_orderdate <= 1998-12-31
  0.00 <= l_discount <= 0.10
  1 <= l_quantity <= 50
  l_returnflag = {R, A, N}
\end{verbatim}
\end{document}